# The quest for the ideal photodetector for the next generation deep underwater neutrino telescopes

B. K. Lubsandorzhiev

*Institute for Nuclear Research of the Russian Academy of Sciences*

*pr-t 60th Anniversary of October, 7A, 117312 Moscow, Russia.*

*Postal address: pr-t 60th Anniversary of October, 7a, 117312 Moscow, Russia; phone: +7-095-1353161; fax: +7-095-1352268;*
*E-mail:* lubsand@pcbai10.inr.ruhep.ru

**Abstract**

We review photodetectors used in present running neutrino telescopes. After a brief historical discourse the photodetector requirements for the next generation deep underwater neutrino telescopes are discussed. It is shown that large area vacuum hybrid phototubes are the closest to the ideal photodetector for such kinds of applications when compared with other vacuum phototubes.

## 1. Introduction.

The history of deep underwater neutrino telescopes spans more than 30 years starting in the middle of 70s of the last century. It was a particular period, full of enthusiasm and hectic work. Neutrinos were opening a new horizon for experimental astrophysics. Due to their neutrality and very weak interaction with matter they can arrive at the Earth undeflected from very distant cosmic source carrying important information about the nature of the source and physics processes in it. High energy neutrinos are of particular importance to shed light on the physics of Active Galactic Nuclei (AGN), binary star systems, gamma ray bursts (GRB) etc. The idea to detect high energy neutrinos was to register Cherenkov light induced by charged leptons and energetic electromagnetic and hadronic showers due to by-products of high energy neutrino interactions with matter. So from the very beginning of intense discussions of the physics capabilities and possible designs of hypothetical neutrino telescopes projects the crucial role of photodetectors has been realized, and much attention has been paid to photodetector developments.

In this paper we will not touch on the photodetector problem in deep under-ice neutrino telescopes like AMANDA and ICECUBE at the South Pole, restricting ourselves just to deep underwater neutrino telescopes.

In the pioneering works of Lescovar, Learned, Bezrukov, Roberts and others [1-4] the main requirements for photodetectors have been thoroughly formulated. But these requirements were based on the knowledge of that time, and in a sense they were in many respects rather naive. The basic ideas of these requirements could be expressed as Olympic principles: *Citius*, *Altius*, *Fortius*. As to photodetectors, the principles were meant to be *faster*, *more sensitive*, and *smarter*. The latter term was introduced in 1983 by G. van Aller and S-O. Flyckt of Philips Laboratories [5]. This term will be defined later in the paper.

A photodetector should have:
a) as high as possible sensitivity to Cherenkov light in water. Because the Cherenkov light spectrum has a $\sim 1/\lambda^2$ behavior, it meant automatically that the photodetector should have high blue sensitivity;
b) large angular acceptance with sensitive area as large as possible, because it was realized since the beginning that planned neutrino detectors will be very sparsely-instrumented arrays to overview huge volumes;
c) time resolution as high as possible, because time resolution will define the angular resolution of the arrays and facilitate background suppression;
d) single photoelectron resolution as high as possible. In other words, the phototube should be as 'smart' as possible, because it was thought that with good enough separation between single and two photoelectron peaks it would be possible to suppress effectively the ocean light background due to $K^{40}$ decay;
e) dark current counting rate as low as possible, because it was thought the level of light background from natural water would be low.

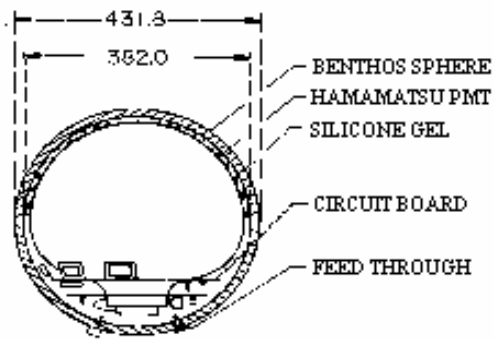

Fig.1. Optical module of DUMAND [7]

The above-listed requirements resulted in the desired design of photodetector for this kind of application – a photodetector with a large hemispherical photocathode to provide a good angular acceptance and probably good isochrony of photoelectrons trajectories, resulting in good time resolution. The photocathode material should be bialkali $K_2CsSb$ for high sensitivity in blue region of the spectrum and have low thermionic noise. The photodetector should be enclosed in a housing to withstand high hydrostatic pressure. The photodetector size is restricted by pressure housing dimension. At that time not much was known of environmental influences on the parameters of the planned arrays or of the photodetector performances required.

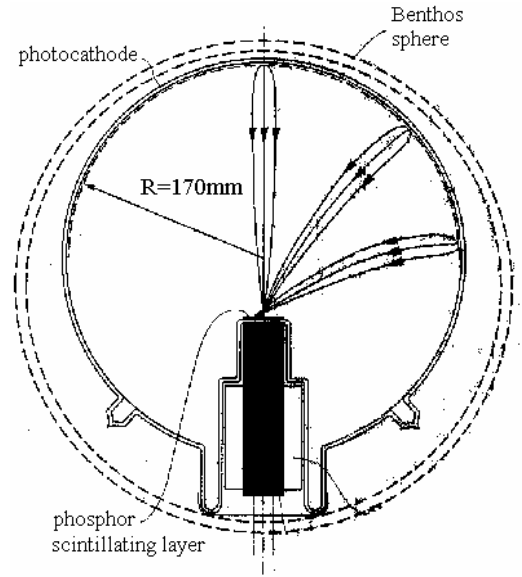

Fig.2. European optical module of DUMAND [9]

The first neutrino telescope project was DUMAND (Deep Underwater Muon and Neutrino Detector) [6]. In that very early effort to implement a neutrino telescope in deep ocean water, two competing tendencies in photodetector development manifested themselves very clearly. The DUMAND 'Japan optical module' (JOM), (Fig. 1), was based on R2018 16" hemispherical classical PMT especially developed for the project by Hamamatsu and named the "DUMAND PMT" [7].

Van Aller, Flyckt and their colleagues from Philips Laboratories at Brive la Gaillarde, France (now Photonis) developed the "Smart" hybrid phototube XP2600 [8]. The 'European optical module' (EOM) [9] based on the XP2600, (Fig. 2), was developed during the final stages of DUMAND project. Unfortunately, DUMAND was terminated and didn't get further development; see [10] for more details.

## 2. Currently running neutrino telescopes.

Despite a lot of work having been done in the frame of deep underwater neutrino telescope development, until recently the only operating deep underwater neutrino telescope was at Lake Baikal. Recently ANTARES entered this esoteric neutrino telescope club [11].

### 2.1. The Lake Baikal neutrino telescope

The lake Baikal neutrino telescope NT-200 [12] is located 3.6 km from the shore and at a depth of 1.1 km. The schematics of NT-200 and the optical modules attached to a detection string are shown in Fig. 3. The telescope consists of 192 optical modules (OMs) grouped pair-wise on 8 vertical strings, which are fixed to an umbrella-like frame. The detector is operated from the shore station by four underwater cables: three electrical cables and one fibre-optic cable. For the latest developments of the experiment see [13].

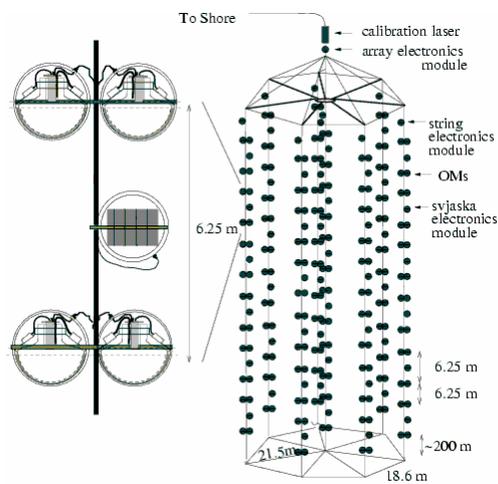

Fig.3. The lake Baikal neutrino telescope NT-200

The optical module of the Lake Baikal experiment is based on the QUASAR-370 a hybrid phototube consisting of an electro-optical preamplifier with a 37 cm hemispherical photocathode and a small conventional type PMT (Fig. 4). A detailed description of the phototube is given in [14-16]. The phototube has been developed especially for neutrino experiments at Lake Baikal and has excellent performance. The QUASAR-370 is equipped with a fast, efficient scintillator ($Y_2SiO_5$:Ce) which provides ~25 pe's in the small PMT for 1 pe from the preamplifier photocathode for 25 kV accelerating electric field. The phototube has on average a single photoelectron resoution of 80% (FWHM) and photoelectron transit time spread (TTS) of 2 ns (FWHM). The QUASAR-370 has no pre-pulses and late pulses, and the level of after-pulses is less than 1%. Pre-pulses, late pulses and after-pulses are inherent phenomena in vacuum phototubes originating from direct photoelectric effect and photoelectron backscattering in a phototube electron multiplication system. We refer the reader to e.g. [17-19] and all references therein for more details. The phototube has ~2000 cm$^2$ sensitive area in $2\pi$ solid angle. Its parameters are immune to the Terrestrial magnetic field. Modified versions of the QUASAR-370 have 1 ns TTS (FWHM) and 35-40% single photoelectron resolution (FWHM) [20-22].

One of the main scientific achievements of the telescope is the discovery of a seasonally varying luminescence in the Lake Baikal deep water [23]. The level of luminescence stays on average at a comparable level to background light found in seawater due to $K^{40}$. Typically the constant level of total counting rate of individual OMs is 50-60 kHz, increasing seasonally to 200-300 kHz.

The deep water optical parameters of the lake are measured continuously. A long-term monitoring of the deep water transparency shows that the maximum transparency is reached at 460-500 nm [24].

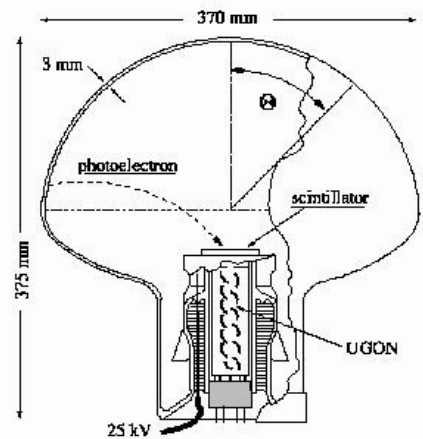

Fig.4. QUASAR-370 phototube

### 2.2. ANTARES neutrino telescope

The ANTARES neutrino telescope [25] is located in the Mediterranean Sea at the depth of ~2400 m and presently consists of 10 strings with 75 OMs on each string. The ANTARES optical module [26], (Fig. 5), is based on 10" PMT of classical type (Hamamatsu R7081-20) with a TTS of 2.7 ns (FWHM) and single photoelectron peak-to-valley ratio of 2-5. The levels of pre-pulses, late pulses and after-pulses are 0.01%, 3.6% and 5% respectively [27]. The PMT has ~500 cm$^2$ sensitive area in a rather narrow solid angle. It is sensitive to the

Terrestrial magnetic field, so it is necessary to use magnetic shields in OMs [27].

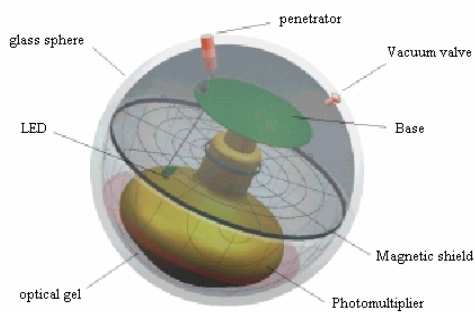

Fig.5. Optical module of the ANTARES neutrino telescope

The site of the ANTARES telescope has a substantially wider "transparency window" in comparison with the Baikal water with a maximum at ~420-520 nm. The counting rate of individual OM is ~50-60 kHz on average, mostly due to $K^{40}$. But sharp short-term increases of counting rates up to 600 kHz-1 MHz due to bioluminescence are also observed [28].

### 3. Influence of water parameters on photodetector requirements

It was shown in [29] how the Cherenkov light spectrum undergoes transformation after passing through 100 m water at Lake Baikal and or in the Mediterranean Sea. The final spectra reaching a photodetector have maxima at 490 nm and 470 nm for Lake Baikal and the Mediterranean Sea respectively. This is very important because the most powerful tool for high energy extraterrestrial neutrino detection is an observation of distant high energy showers produced by neutrino interactions in water. In this respect it would be better for photodetectors of future deep underwater neutrino telescopes to have extended green sensitive bi- or multi-alkali photocathodes. As was mentioned in the previous sections, OM counting rates are dominated by natural water light background, so the requirement on the photodetector's cathode thermionic noise is fairly loose. Light dispersion in natural deep water plays a conspicuous role too [30, 31]. As was demonstrated in [29] at distances of 100 m or more from a light source to a photodetector in the Mediterranean Sea a light pulse's width widens by ~5 ns due to dispersion, so the requirement for photodetector time resolution is correspondingly also loosened.

### 3. Conclusion

Experimental high energy neutrino astrophysics is entering its maturing stage. Currently two deep underwater neutrino telescopes are successfully operating: the Lake Baikal neutrino telescope and the ANTARES detector in the Mediterranean Sea. But unfortunately their effective volumes are relatively small. It is very likely that even 1 $km^3$ arrays will be too small to detect very high energy extraterrestrial neutrinos with good statistics. So, new projects for giant, deep underwater Cherenkov arrays are already presented a challenge in the development of experimental technique.

The development of deep underwater neutrino telescopes has pushed forward the development of photodetectors. The experience accumulated during the development and operation of the present day neutrino telescopes allows formulation of new requirements for photodetectors and demonstrates that large area hybrid phototubes like the QUASAR-370 are rather close to the ideal photodetector for the next generation giant deep underwater neutrino telescopes; possibly serving as a prototype for the development of a new photodetector capable to match the challenge.

The author is grateful to Dr. V. Ch. Lubsandorzhieva for many useful discussions and invaluable remarks.